\documentclass[12pt]{PoS}

\usepackage{graphicx}
\usepackage{dcolumn}
\usepackage{latexsym}
\usepackage{dsfont}
\usepackage{bm}

\newcommand{\ov}[1]{\overline{#1}}
\newcommand{\be}{\begin{equation}}
\newcommand{\ee}{\end{equation}}
\newcommand{\bey}{\begin{eqnarray}}
\newcommand{\eey}{\end{eqnarray}}

\title{Fermion loop simulations in 2--d lattice theories -- results and
 limitations}

\ShortTitle{Fermion loop simulation}

\author{\speaker{Markus Limmer}\\
        Institut f\"ur Physik, FB Theoretische Physik,
        Universit\"at Graz, 8010 Graz, Austria\\
        E--mail: \email{markus.limmer@uni-graz.at}}

\author{Christof Gattringer\\
        Institut f\"ur Physik, FB Theoretische Physik,
        Universit\"at Graz, 8010 Graz, Austria\\
        E--mail: \email{christof.gattringer@uni-graz.at}}
        
\author{Verena Hermann\\
        Department of Earth and Environmental Sciences, Geophysics,
        Munich University, 80333 Munich, Germany\\
        E--mail: \email{verena.hermann@geophysik.uni-muenchen.de}}

\abstract{We review our results for the simulation of the 2d lattice 
Gross--Neveu model in a fermion loop representation. Possible extensions of
our techniques to other models and higher dimensions are discussed, as well as
the limitations of loop--type representations.}

\FullConference{The XXV International Symposium on Lattice Field Theory\\
		 July 30 - August 4, 2007\\Regensburg, Germany}

\begin{document}

\section{Introduction}

\noindent 
Numerical simulations of fermion systems have to deal with the Pauli
principle which enforces a completely anti--symmetric wave function for
fermions. Obviously, this requirement makes fermion systems extremely
non--local. Changing the system at a single point affects all degrees of
freedom. In more technical terms, an ad--hoc local change may lead to a
completely different value of the fermion determinant. Thus intricate methods
such as the hybrid Monte Carlo algorithm were developed. Nevertheless, also
these methods have to deal with the non--locality which makes
the simulation of fermions several orders of magnitude more expensive than
bosonic systems. 

Circumventing the non--locality problem altogether is certainly an extremely
appealing idea. A prominent example of a breakthrough in this direction is the 
Meron Cluster Algorithm proposed in \cite{wiese}, which allows for highly
effective Monte Carlo simulation for certain classes of fermionic models. 

An alternative approach is a formal solution of the Grassmann path integral
for fermions which represents the partition function of the system as a model
of closed fermion loops. While for a long time this technique
has been known to work well in the strong coupling
limit, only recently \cite{gattringer_gn,gattringer_8v}
loop representations were found for 
two--dimensional lattice field theories at arbitrary coupling,
in particular the Gross--Neveu model \cite{gn}.

In a recent publication
\cite{grazloop} it was shown that the loop representation allows for an
efficient and considerably cheaper simulation than traditional methods. 
In a subsequent paper \cite{wolff} Wolff has rederived the loop representation
by decomposing 2--d Dirac fermions into Majorana components and
demonstrated that the loop
formulation can be recast as a spin system where a cluster algorithm boosts
the efficiency of a numerical simulation further.

In this contribution we review the loop representation of the lattice
Gross--Neveu model and its use for a numerical simulation. We furthermore
discuss the loop representation of the Schwinger Model \cite{sm}, i.e., QED in
two dimensions, an example 
which illustrates the limitations of the loop approach for a
use in numerical simulations. 
  
\section{Loop representation of the lattice Gross--Neveu model}

\noindent
We consider the lattice Gross--Neveu model with $N$ flavors of Wilson
fermions. The lattice action for the fermions is given by
(we set the lattice spacing to $a=1$)
\begin{eqnarray}
S_F[\overline{\psi},\psi,\varphi] & \; = \; &
\sum_{x \in \Lambda} \ov{\psi}(x) \,D(x,y) \psi(y) \; ,
\nonumber 
\\
D(x,y) & \; = \; & \big[ \, 2 \, + \, m \, + \, \varphi(x) \, \big] \, \delta_{x,y} \;
- \; \sum_{\mu = \pm 1}^{\pm2} \frac{1 \mp \gamma_\mu}{2} \, 
\delta_{x+\hat{\mu},y} \; .
\label{fermionaction}
\end{eqnarray}
The sum runs over the 2--d lattice $\Lambda$. In two dimensions the
 $\gamma$--matrices may
be chosen as the Pauli matrices, $\gamma_{\pm \mu} = \sigma_\mu$.
The spinors $\overline{\psi}$ and $\psi$ are vectors of $N$ 2--d
spinors, and we use vector/matrix notation for both the spinor and flavor
indices. Through the Dirac operator $D(x,y)$ all flavors couple in the same 
way to the real scalar field $\varphi$, which has the action 
\begin{equation}
S_S[\varphi] \; = \; \frac{1}{2g} \sum_{x \in \Lambda} \varphi(x)^2 \; .
\label{scalaraction}
\end{equation}
When integrating out the scalar field the 4--fermi interaction
\begin{equation}
- \, \frac{g}{2} \, \sum_{x \in \Lambda} \Big( \ov{\psi}(x) \psi(x) \Big)^2 
\end{equation}
is induced. The partition function of the model is given by
\be
Z \; = \; \int \prod_x d \varphi(x) \, d \ov{\psi}(x)
\, d \psi(x) \, e^{  - \, S_S[\varphi] \, - \, 
S_F[\ov{\psi}, \psi, \varphi ]} 
\; = \; \int \prod_x d \varphi(x) \, e^{  - \, S_S[\varphi]} \,
\det\big( D[\varphi] \big)^N \; ,
\label{partitionfunction}
\ee
where in the second step the fermions were integrated out giving rise to a
remaining path integral over the scalar field with the fermion determinant
raised to the power $N$ as integrand. 

The last expression is a possible starting point for identifying the loop 
representation. The Dirac operator $D$ may be rewritten as
\begin{eqnarray}
D(x,y) & \; = \; & \big[ \, 2 \, + \, m \, + \, \varphi(x) \, \big] \, 
\Big[ \, \delta_{x,y} \; - \; H(x,y) \, \Big] \; ,
\label{hoppingform}
\\
H(x,y) & \; = \; & \frac{1}{2 \, + \, m \, + \, \varphi(x)} \,
\sum_{\mu = \pm 1}^{\pm2} \frac{1 \mp \gamma_\mu}{2} \, 
\delta_{x+\hat{\mu},y} \; ,
\nonumber
\end{eqnarray}
where we have combined all nearest neighbor terms in the hopping matrix $H$. 
Inserting the representation (\ref{hoppingform}) into the partition function 
(\ref{partitionfunction}) one finds
\begin{eqnarray}
Z & \; = \; & \int \prod_x d \phi(x) \, e^{  - \, S_S[\varphi]} \,
\prod_x \, \big( \, 2 \, + \, m \, + \, \varphi(x) \, \big)^{2N} \, 
\det\big( 1 - H[\varphi] \big)^N 
\nonumber \\
& \; = \; & \int \prod_x d \phi(x) \, e^{  - \, S_S[\varphi]} \,
\prod_x \, \big( \, 2 \, + \, m \, + \, \varphi(x) \, \big)^{2N} \,
\left( \exp \left( - \sum_{n=1}^\infty \frac{1}{n} \, \mbox{Tr} \; 
\Big[ H^n \Big]
\right) \right)^N \; ,
\label{hoppingexpansion}
\end{eqnarray}
where we have used the formula $\det[1-H] = 
\exp(\mbox{Tr} \; \ln [1-H])$ for the determinant and expanded the logarithm.

The expression (\ref{hoppingexpansion}) is the well known hopping
expansion. At this point the loops are already evident: The hopping matrix $H$
is a matrix which describes hopping between neighboring lattice
points. Consequently the power $H^n$ in (\ref{hoppingexpansion}) corresponds
to a chain of $n$ subsequent steps. When taking the trace only closed chains,
i.e., loops survive. Such an expansion holds in arbitrary dimensions and for
different types of bosonic fields, scalar, as well as gauge fields. 

The crucial step, however, is that the traces $\mbox{Tr} \; [ H^n ]$ in
(\ref{hoppingexpansion}) can be evaluated only in special cases. In addition
to the space--time indices, this trace is over Dirac and for non--abelian gauge
theories also over the color indices. For the latter a simple closed
form is probably not realistic. 
Concerning the Dirac indices, in two dimensions 
it is possible \cite{nucu} to find a closed form for the trace 
over the matrices 
$[1 \pm \gamma_\mu]/2$ which enter the hopping matrix (\ref{hoppingform}).
Thus for the case of non--abelian interactions in two dimensions the exponent
in (\ref{hoppingexpansion}) can be computed in closed form. 

As discussed, for special cases (scalar or abelian bosonic fields in 2--d)
the coefficients for the individual loops in
the exponent of (\ref{hoppingexpansion}) can be computed analytically. The
final step is to bring the loops down from the exponent. Here two different
approaches were followed in \cite{gattringer_8v} and \cite{sm}. In the former
case the final expression for the loop representation was obtained by
comparing the 2--d Wilson fermions to the hopping expansion of a 8--vertex
model. In the latter case a direct evaluation of the exponential of the sum
over loops was performed. Again we remark that there is also the direct
identification of the loop representation through the explicit solution of 
the Grassmann integral for the Majorana components \cite{wolff}. 

Once the determinant is given as a sum over loops with known coefficients
(not as the exponential of a sum over loops) there is only the path integration
over the bosonic variables attached to the loops left to be done. For the case
of the scalar fields which give rise to the 4--fermi interaction, the path
integration is trivial, since at each lattice point only moments of the 
Gaussian distribution need to be computed. 
In this way the partition function of the $N$--flavor lattice
Gross--Neveu model is found to be a model of $2N$ self--avoiding loops. For the
case of general $N$ we refer the reader to \cite{gattringer_gn}, and here
quote the result for $N=1$, which is the case that was used in the numerical 
simulations \cite{grazloop,wolff}. The partition function reads

\be
Z \; = \; \sum_{r,b} \, \left( \frac{1}{\sqrt{2}} \right)^{c(r,b)} \, 
f_1^{\,n_1(r,b)} \, f_2^{\,n_2(r,b)} \ .
\label{zloop}
\ee
The sum runs over two sets of loops which we refer to as red
($r$) and blue ($b$). For a given color the loops are self avoiding, i.e.,
they cannot cross or touch each other, while loops of different may do so. 
In Eq.~(\ref{zloop}) $c(r,b)$ is the total number of corners for both,
red and blue loops. Thus every corner contributes a factor of 
$1/\sqrt{2}$ to the weight of a configuration.
Furthermore, 
$n_1(r,b)$ is the number of lattice sites which 
are singly occupied by either $r$ or $b$ and 
$n_2(r,b)$ is the number of doubly occupied sites, i.e., sites which are
visited by both, a red and a blue loop. 
The weight factors $f_1$ and $f_2$ are simple functions,
related to the mass $m$ and the coupling $g$ through
\be
f_1 \, = \, \frac{2+m}{(2+m)^2 + g} \quad , \quad f_2 \; = \; 
\frac{1}{(2+m)^2 + g} \ .
\label{f1f2}
\ee
The mapping (\ref{zloop}), (\ref{f1f2}) is exact in the thermodynamic
limit. For finite volume different types of boundary conditions in the two 
representations lead to finite size effects: In the loop 
representation we need to have closed loops and in a finite volume
the loops can wind around the periodic boundary. The loop configurations
fall into three equivalence classes, $C^{ee}, C^{eo}, C^{oo}$, depending on 
the numbers of red and blue non--trivially winding loops 
(see also \cite{galasa}): 
$C^{ee}$ (even--even): The total number of windings for both,
red and blue loops is even for both directions. $C^{eo}$ (even--odd):
One of the colors has an 
odd number of windings for one of the directions. $C^{oo}$ (odd--odd):
Both colors
have an odd number of windings in one of the directions. These 
equivalence classes cannot be linked in a simple way to the boundary 
conditions in the standard representation which we discussed above.
However, in \cite{grazloop} it was shown that the boundary effects vanish as
$1/\sqrt{V}$, with $V$ denoting the volume. The representation in terms of the 
Ising spin variables \cite{wolff} solves the boundary condition problem
completely, and the partition functions of the original fermionic-- and the
spin representation are identical also on finite volumes. 

\begin{figure}
\begin{center}
\includegraphics[width=0.49\textwidth,clip]{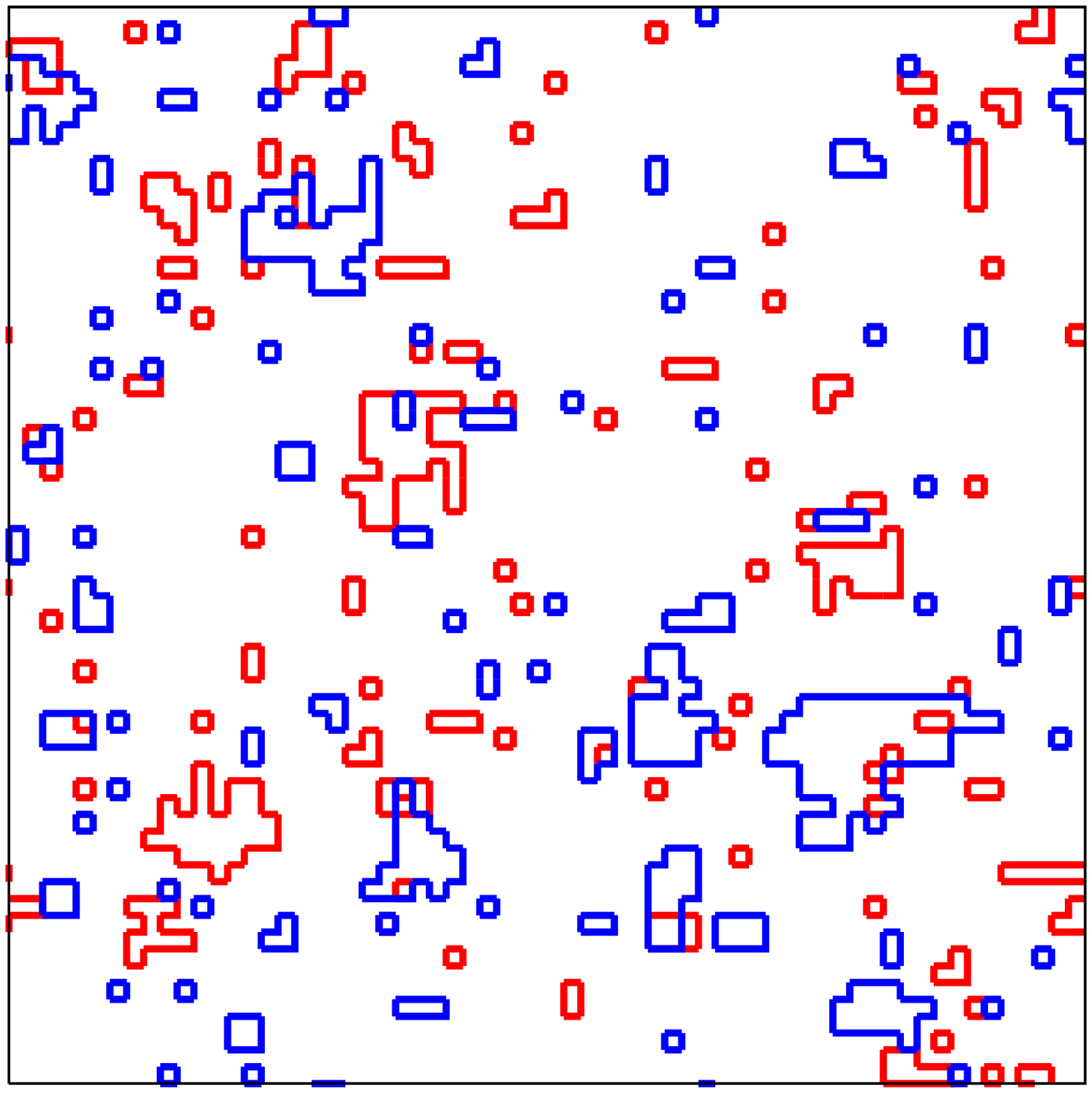} \hfill
\includegraphics[width=0.49\textwidth,clip]{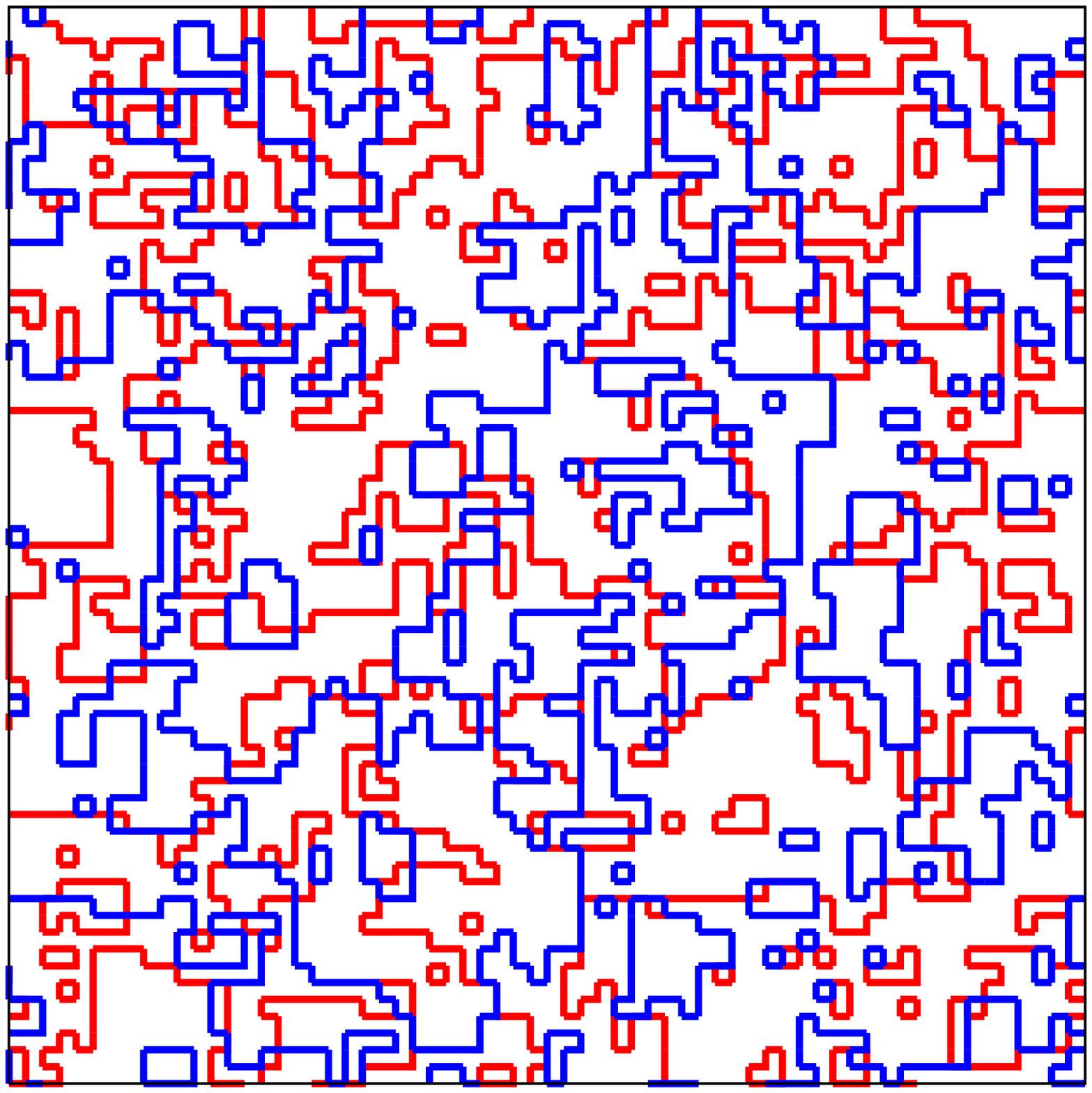} 
\end{center}
\caption{Typical loop configurations in the 1--flavor Gross--Neveu model.
We compare two different values of the parameters. L.h.s.: $g = 0.0, m = 0.1$;
r.h.s.: $g=0.0, m = 0.2$.} 
\label{fig1}
\end{figure}

\section{Numerical simulation}

For the numerical simulation of the loop representation 
of the Gross--Neveu model
we use a local Metropolis update. Red and blue
loops are updated alternately by performing a full 
sweep through the lattice for
only one color and meanwhile treating 
the other as a constant background field.
During one sweep all plaquettes are visited once. 
A trial configuration is offered
by inverting the 4 links of the current plaquette 
of the active color. With such an
offer, we guarantee that the loops stay closed, or 
new loops are created if all links
of the plaquette were empty before. In case that 
the self--avoidance condition is
violated, the proposal is rejected. Otherwise
the new configuration is accepted with 
the Metropolis probability
\be
p \, = \, \left(\frac{1}{\sqrt{2}}\right)^{\Delta c}\, 
f_1^{\Delta n_1}\, f_1^{\Delta n_1}\ .
\ee
$\Delta c$ is the difference of the number of corners, 
$\Delta n_1$ and $\Delta n_2$ are
the differences in the occupation numbers.
Fig.~\ref{fig1} shows snapshots of typical loop configurations in the
numerical simulation. 

Particularly simple 
observables are derivatives of the free energy $F=-\ln Z$. These expressions
can be written as moments of occupation numbers. To be more explicit
we discuss 
the chiral condensate $\chi$ and its susceptibility $C_\chi$. 
The conventional definitions are
\bey
\chi & = & \frac{1}{V} \sum_{x\in\Lambda} \langle \ov{\psi}(x)\psi(x) \rangle
     \  = \  -\frac{1}{V} \frac{\partial \ln Z}{\partial m}\ ,
\\
C_\chi & = & \frac{\partial \chi}{\partial m}\ .
\eey
In terms of loop variables these expressions read
\bey
\chi &=& -\, \frac{1}{V f_1}\ \big[ f_2 \langle n_1 \rangle
\, + \, 2 f_1^2 \langle n_0 \rangle \big] \ , \label{occobs1}
\\
C_\chi &=&  -\, \frac{1}{V f_1^2}\ \big[ (4f_1^4 - 2f_1^2 f_2)\, 
\big\langle (n_0 - \langle n_0 \rangle )^2 \big\rangle 
\, + \, (f_2^2 - 2 f_1^2 f_2) 
\big\langle (n_1 - \langle n_1 \rangle )^2 \big\rangle 
\nonumber \\
& & + \, 2 f_1^2f_2 \, 
\big\langle (n_0 + n_1 - \langle n_0 + n_1 \rangle)^2 \big\rangle
\, - \, (4f_1^4 - 2f_1^2 f_2) \langle n_0 \rangle - f_2^2 \,
\langle n_1 \rangle \big] \; , \label{occobs2}
\eey
where the number $n_0$ is the total number of empty lattice sites. These
representations were obtained by differentiating the partition function 
(\ref{zloop}).

\begin{figure}
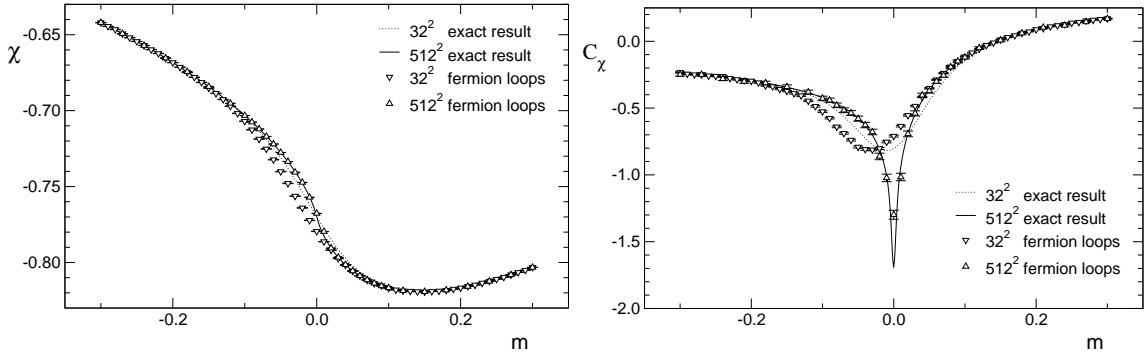

\begin{center}
\includegraphics[width=0.495\textwidth,clip]{chi.eps} \hfill
\includegraphics[width=0.495\textwidth,clip]{chisusz.eps} 
\end{center}
\caption{L.h.s.: The chiral condensate $\chi$ for $g=0$ as a function of $m$
for 2 different lattice sizes. We compare the simulation 
in the loop representation (symbols with error bars) to the exact 
result from Fourier transformation (curves). R.h.s.: Same as on the l.h.s.,
now for the chiral susceptibility $C_\chi$.}
\label{fig2}
\end{figure}

In Fig.~\ref{fig2} we compare the loop results at $g=0$ (symbols) with
those from Fourier transformation (curves) which is applicable in standard
representation for the special case of vanishing coupling. Two volumes are used, 
the relatively small lattice $32\times 32$ and a substantially larger lattice 
of $512 \times 512$. For each value of $m$ typically $10000$ sweeps were 
used to equilibrate the system and then about $50000$ measurements were 
performed for our observables. These measurements have been separated
by $10$ sweeps for each color to guarantee sufficient decorrelation. 
For the calculation of the observables we used Eqs.~(\ref{occobs1}), 
(\ref{occobs2}), and the statistical error
was computed with the jackknife method. For the larger lattice we find almost
perfect agreement of the results from the loop representation with the
analytic results. In \cite{grazloop} we have presented the results from the
loop simulation for several values of $g \neq 0$ and we compared these to the
outcome of a simulation with traditional techniques. Concerning the
performance we find that with comparable amounts of CPU time with the loop
representation we can work on volumes which are two orders of magnitude larger
than those available with traditional techniques. The cluster approach of 
\cite{wolff} enhances this performance further. 

\section{Perspectives and limitations of the loop representation}

Having addressed the merits of the loop representation for the Gross--Neveu
model, we would like to comment on possible extensions of the loop method, but
also discuss the points where we see limits of the method. 

We begin this discussion with stressing that, although we so far 
restricted our numerical simulations to only two flavors, a generalization to 
the Gross--Neveu model with an arbitrary number of flavors is straightforward
with the loop formula given in \cite{gattringer_gn}. 

Concerning models with relativistic fermions which are coupled via 4--fermi
interactions in higher dimensions, one could try
to repeat the strategy that led to the loop representation for the 2--d models.
An essential step in the identification of the loop formalism was the
closed result for the traces of the $\gamma$--matrices. While this is a
relatively simple problem in 2--d, the corresponding structures in 4--d are
considerably more involved. For an attempt to find such a closed formula in
four dimensions see, e.g., \cite{Scharnhorst}.   

Interesting might also be the case of non--relativistic fermions in
2+1 dimensions with 4--fermi interaction. For some of these systems a relation to
3--d spin models is known \cite{samuel,surface} which might be useful for
a numerical simulation.  

We finally comment on the applicability of the loop approach to lattice gauge
theories beyond the strong coupling limit. We have already mentioned, that we
judge the case of non--abelian gauge fields as an elusive goal, due to the
non--commutativity of the link variables. For abelian gauge fields the
situation is simpler and in 2--dimensions the loops in (\ref{hoppingexpansion})
can again be computed in closed form. The resulting loop representation for
the lattice Schwinger Model \cite{sm} is of a different type, however. Since
gauge fields are oriented quantities, one has to work with oriented loops,
while the loops for a scalar interaction are non--oriented (see
Eq.~(\ref{zloop})). One finds that reverting the 
orientation of a loop corresponds to complex conjugation of its contribution. This
implies, that certain cancellations among loops, which simplify the scalar
case, are no longer possible \cite{sm}. The loops for the Schwinger Model 
turn out to be self--intersecting and an extra minus sign appears for each
intersection. In a numerical simulation \cite{steiner} it was found that the
resulting fermion sign problem limits the size of the accessible volumes. At
the moment it is unclear whether this is a fundamental obstacle or if this
problem can be overcome by different techniques.

{\bf Acknowledgments:} 
We thank Erek Bilgici, Philipp Huber, Christian Lang, Klaus Richter, Andreas
Sch\"afer and Erhard Seiler for discussions and helpful remarks. This work was
supported by \textit{Fonds zur F\"orderung der wissenschaftlichen Forschung
in \"Osterreich (FWF DK W1203--N08)}. The simulations were done at
the ZID cluster of the Karl--Franzens Unversity Graz.

\end{document}